\documentclass[5p]{elsarticle}
\usepackage{lineno,hyperref}
\modulolinenumbers[5]
\usepackage{graphicx}   
\usepackage{latexsym}   
\usepackage{bm}

\newcommand{\goo}{\,\raisebox{-.5ex}{$\stackrel{>}{\scriptstyle\sim}$}\,}
\newcommand{\loo}{\,\raisebox{-.5ex}{$\stackrel{<}{\scriptstyle\sim}$}\,}

\begin{document}
\begin{frontmatter}

\title{Formation of hypermatter and hypernuclei within transport models 
in relativistic ion collisions.}

\author{A.S.~Botvina$^{1,2}$, J.~Steinheimer$^{1}$, E.~Bratkovskaya$^{1}$, 
M.~Bleicher$^{1}$, J.~Pochodzalla$^{3,4}$}


\address{$^1$Frankfurt Institute for Advanced Studies, J.W.Goethe 
University, D-60438 Frankfurt am Main, Germany} 
\address{$^2$Institute for Nuclear 
Research, Russian Academy of Sciences, 117312 Moscow, Russia} 
\address{$^3$Helmholtz-Institut Mainz, J.Gutenberg-Universit{\"a}t, 
55099 Mainz, Germany} 
\address{$^4$ Institut f{\"u}r Kernphysik and PRISMA Cluster of 
Excellence, J.Gutenberg-Universit{\"a}t Mainz, D-55099 Germany}

\date{\today}

\begin{abstract}
Within a combined approach we investigate the main features of the 
production of hyper-fragments in relativistic heavy-ion collisions. 
The formation of hyperons is modelled within the UrQMD and HSD transport 
codes. To describe the hyperon capture by nucleons and nuclear residues 
a coalescence of baryons (CB) model was developed. 
We demonstrate that the origin of hypernuclei of various masses can be 
explained by typical baryon interactions, and that it is similar to processes 
leading to the production 
of conventional nuclei. At high beam energies we predict a saturation of the 
yields of all hyper-fragments, therefore, this kind of reactions can be 
studied with high yields even at the accelerators of moderate relativistic 
energies. 
\end{abstract}


\end{frontmatter}



\section{Introduction}

The investigation of hypernuclei is a rapidly progressing field of nuclear 
physics, since these nuclei provide complementary methods to improve 
traditional nuclear studies and open new horizons for studying nuclear 
physics aspects related to particle 
physics and nuclear astrophysics (see, e.g., 
\cite{Ban90,Sch93,Gre96,Has06,Sch08,Gal12,Buy13,Hel14} 
and references therein). 
Indeed, baryons with strangeness embedded in a nuclear environment are 
the only 
available tool to approach the many-body aspects of the three-flavor 
strong interaction. 
Hyperon-nucleon and hyperon-hyperon interactions are also an essential 
ingredient for the nuclear Equation of State (EOS) at high density and 
low temperature. 
Another novel aspect of contemporary hypernuclear studies is the exploration 
of the limits of stability in isospin and strangeness space.

Presently, hypernuclear physics is still focused on spectroscopic 
information and is dominated by a quite limited set of reactions. These 
are reactions induced by high-energy hadrons and leptons leading to the 
production of only few particles, including kaons which are often used to 
tag the production of hypernuclei in their ground and low excited 
states. In such reactions hyper-systems with baryon density around the 
nuclear saturation density, $\rho_0 \approx 0.15$ fm$^{-3}$ are formed. 
Therefore, most previous theoretical studies concentrated on the 
calculation of the structure of nearly cold hypernuclei.
However, many experimental collaborations (e.g., PANDA \cite{panda}, 
FOPI/CBM, and Super-FRS/NUSTAR at FAIR \cite{super-frs,Rap13}; 
STAR at RHIC \cite{star}; ALICE at LHC \cite{alice}; BM@N and MPD at 
NICA \cite{nica}) have started or plan to investigate hypernuclei and 
their properties in hadron and heavy ion induced reactions. This represents 
an essential extension of nuclear/hypernuclear studies: 
The isospin space, particle unstable states, multiple strange 
nuclei, the production of hypermatter, and precision lifetime measurements 
are unique topics of these fragmentation reactions.

It is relevant in this respect to note that the very first experimental 
observation of a hypernucleus was obtained in the 1950-s in nuclear 
multifragmentation reactions induced by cosmic rays \cite{Dan53}. 
In recent years a remarkable progress was made in the investigation of the 
fragmentation and multifragmentation reactions associated with relativistic 
heavy-ion collisions (see, e.g., \cite{Bon95,Xi97,Sch01,Ogu11,Gos97} 
and references therein). This gives us an opportunity to apply well known 
theoretical methods, which were adopted for the description of the 
conventional reactions, also for the formation of hypernuclei 
\cite{Bot07,Das09}.
The task of this work is to develop new realistic models of hypernuclear 
production which are able to provide detailed predictions in order 
to optimise the experimental conditions when searching for both 
$\Lambda$-hypernuclei and normal exotic nuclei.

The formation processes of hypernuclei are quite different in central 
and peripheral ion collisions. There are indications that in high energetic 
central 
collisions the coalescence mechanism, which assemble 
light hyper-fragments from the produced hyperons and 
nucleons (including anti-baryons) is dominating 
\cite{star,alice,ygma-nufra,camerini-nufra}.
Because of the very high temperature of the fireball (T$\approx$160 MeV) 
only lightest clusters, with mass numbers A$\loo$4, can be 
produced in this way with a reasonable yield \cite{Ste12}. 
On the other hand, it was noticed sometime ago that the capture of 
hyperons in spectator regions 
after peripheral nuclear collisions is a promising way to produce 
hypernuclei \cite{Wak88,Cas95,giessen,Bot11}. Nuclear matter created 
in peripheral collisions shows distinctly different
properties compared to nuclear matter at mid-rapidity. 
It is well established that moderately excited spectator residues 
(T$\loo$5-6 MeV) are produced in such reactions 
\cite{Bon95,Xi97,Sch01,Poc97}. A hyperon bound 
in these residues should not change the picture since the 
hyperon-nucleon forces are of the same order as the nucleon-nucleon 
ones. General features of the decay of such hyper-residues into 
hyper-fragments could be investigated with statistical models 
(e.g.,generalized Statistical Multifragmentation Model SMM 
\cite{Buy13,Bot07}), which successfully describe the production of 
normal fragments 
\cite{Bon95,Xi97,Sch01,Ogu11}. The models predict the formation of 
exotic hypernuclei and hypernuclei beyond the drip-lines, which are difficult 
to create in other reactions \cite{Buy13}. There is an 
alternative treatment of the process that considers first statistical 
SMM decay of excited residues, and, afterwards, a coalescence model 
for final production of hyper-fragments \cite{giessen}. Both theoretical 
mechanisms are under discussion and waiting for a test by experiments.
Spectator heavy fissioning hypernuclei were identified with a 
relatively high probability in reactions induced by protons with energy 
around the threshold \cite{Ohm97}, and in annihilation of antiprotons 
\cite{Arm93}. Very encouraging results on hypernuclei 
come from experiments with light projectiles: In addition to well-known 
hypernuclei \cite{saito-new}, evidences for unexpected exotic hypernuclear 
states, like a $\Lambda$ hyperon bound to two neutrons, were reported 
\cite{Rap13b}, which were never observed in other reactions. As was 
discussed, the production of such new exotic states could be naturally 
explained 
within the break-up of excited hypernuclear systems \cite{Rap13b,Bot12}.

In previous publications we have considered the formation of hypernuclei 
within the Dubna cascade model (DCM) \cite{Ton83,toneev90} and the 
Ultra-relativistic Quantum Molecular Dynamics model (UrQMD) \cite{Bas98}. 
These calculations include the capture of the produced hyperons in the 
potential 
of the spectator residues \cite{Bot11,Bot13}, and the coalescence into 
lightest clusters together with their thermal production in central 
collisions \cite{Ste12}. Involving new transport models is very 
important since we obtain knowledge about uncertainties in such 
calculations. In this work, besides UrQMD, we employ the hadron-string 
dynamics (HSD) model \cite{Bra08}, which were used successfully for 
description of strangeness production \cite{Bra04,Har12}. We develop a 
generalization of the coalescence model \cite{Neu00}, the coalescence of 
baryons (CB), which is applied after UrQMD and HSD stage. In this way it is 
possible to form 
fragments of all sizes, from the lightest nuclei to the heavy residues, 
including hypernuclei within the same mechanism.  
The advantage of this procedure is the possibility 
to predict the correlations of yields of hypernuclei, including their sizes, 
with the rapidity on the 
event-by-event basis, that is very essential for the planning of future 
experiments.


\section{Transport calculations of conventional and strange baryons}


A detailed picture of peripheral relativistic heavy-ion collisions has 
been established in many experimental and theoretical studies. 
Nucleons from the overlapping parts of the projectile and target 
(participant zone) interact intensively between each other 
and with other hadrons produced in primary and secondary collisions. 
Nucleons from the non-overlapping parts interact rarely, and 
they form the residual nuclear systems, which we call spectators. 
We apply two dynamical models to describe the processes leading to 
the production of strange particles in nucleus-nucleus collisions 
before their accumulation in nuclear matter. Using different models allows us 
to estimate the theoretical uncertainties associated with the different 
treatment of the dynamical stage. 

The first model is the Ultra-relativistic Quantum Molecular Dynamics model 
(UrQMD)~\cite{Bas98,Ble99}. 
The model is based on an effective microscopic solution of the 
relativistic Boltzmann equation. Products of binary interactions of 
particles include 39 different hadronic 
species (and their anti-particles) which scatter according to their 
geometrical cross section. The allowed processes include elastic 
scattering and $2\rightarrow n$ processes via resonance creation (and 
decays) as well as string excitations for large center-of-mass energies 
($\sqrt{s}\goo 3$ GeV) . The current version 3.4 of UrQMD also 
includes important strangeness exchange reactions, e.g., $\overline{K}+N 
\leftrightarrow \pi +Y$ (where $Y$ is a strange baryon) \cite{sxch}. 


Another model is the off-shell Hadron-String-Dynamics (HSD) transport 
model \cite{Bra08,Cas99}. It is based on the solution of the 
generalized transport equation \cite{Cas00} including covariant self 
energies for the baryons. 
We recall that in the HSD approach nucleons, $\Delta$'s, 
N$^*$(1440), N$^*$(1535), $\Lambda$, $\Sigma$ and $\Sigma^*$ 
hyperons, $\Xi$'s, $\Xi^*$'s and $\Omega$'s as well as their 
antiparticles are included on the baryonic side, whereas the $0^-$ 
and $1^-$ octet states are incorporated in the mesonic sector. 
Inelastic baryon--baryon (and meson-baryon) collisions with energies 
above $\sqrt{s_{th}}\simeq 2.6$ GeV (and  $\sqrt s_{th}\simeq 2.3$ GeV, 
respectively) 
are described by the FRITIOF string model \cite{And93}, whereas low 
energy hadron--hadron collisions are modelled using experimental 
cross sections. 

In the both HSD and UrQMD models the initial state of colliding nuclei 
is generated 
similarly: The nucleon's coordinates are initialized according to a 
Woods-Saxon profile in coordinate space and their momenta are assigned 
randomly according to the Fermi distribution.


\section{Coalescence of baryons} 

A composite particle can be formed from two or more nucleons if they are 
close to each other in phase space. This simple prescription is known 
as coalescence model and it is based on the properties of the 
nucleon--nucleon interaction. One can use the coalescence in both momentum 
(velocity) space and the coordinate space. 
The coalescence in the momentum space model has proven successful in 
reproducing experimental data on the production of light clusters 
(see e.g. \cite{Ste12,Ton83}).

Recently, we developed an alternative formulation of the coalescence model, 
the 
coalescence of baryons (CB), which is suitable for computer event by 
event simulations \cite{Neu00}. 
Baryons (nucleons and hyperons) can produce a cluster with mass number $A$ 
if their velocities relative to the center-of-mass velocity of the cluster 
is less than $v_c$. Accordingly we require 
$|\vec{v}_{i}-\vec{v}_{cm}|<v_{c}$ for all $i=1,...,A$, where 
$\vec{v}_{cm}=\frac{1}{E_A}\sum_{i=1}^{A}\vec{p}_{i}$ ($\vec{p}_{i}$ are 
momenta and $E_A$ is the sum energy of the baryons in the cluster). 
This is performed by sequential comparison of the velocities of all baryons. 

If we consider only the production of lightest clusters (A$\loo$4) the 
coalescence velocity parameter $v_{c}\approx 0.1 c$ gives a good description 
of the data, as was shown in previous analyses \cite{Ste12,Ton83}. 
However, the coalescence mechanism may also be applied to construct 
heavy nuclei \cite{Neu00}. 
In this case the parameter $v_{c}$ should be larger, in order to 
incorporate higher velocities of the hyperons which can be captured in the 
deeper potentials of big nuclei. This potential well saturates at 
around $\sim$30--40 MeV. It was demonstrated in Ref.~\cite{Bot11} (fig.~10) 
that according to this potential criterion the momentum distribution of the 
captured $\Lambda$ hyperons can be approximated by a step-like function, and 
that hyperons with relative momenta less than 200-250 MeV/c can be bound. 
Therefore, relative velocities up to $0.25 c$ should be taken into account as 
coalescence parameter, which is naturally close to the Fermi velocity. 

We would like to note a problem which is sometimes disregarded in coalescence 
simulations. Some nucleons may have velocities such that they can belong to 
different (or even more than two) coalescent clusters according to the 
coalescence criterion. 
In these cases the final yield will depend on the sequence of nucleons 
within the algorithm. To avoid this uncertainty we developed an iterative 
coalescence procedure. $M$ steps are calculated in the coalescence 
routine with the radius $v_{cj}$ which is increased at each step $j$: 
$v_{cj}=(j/M)\cdot v_c$ (with $j=1,...,M$). Clusters produced at earlier 
steps participate as a whole in the following steps. In this case the final 
clusters not only meet the coalescence criterion but also their nucleons 
have the minimum distance in the velocity space. This procedure gives a 
mathematically correct result in the limit $M\rightarrow\infty$, however, 
we found that in practical calculations it is sufficient to confine the steps 
to $M$=5. 

For more reliable identification of the clusters, we apply in addition a 
coordinate 
proximity criterion. A single baryon and a cluster with mass number A 
is confirmed to compose the new cluster if the relative distances of all 
baryons from the cluster's center of mass $r_c$ is less than 
$r_0\cdot$(A+1)$^{1/3}$. Here is $r_0$=2~fm, which can be justified by 
multifragmentation studies (see, e.g., \cite{Bon95}): It was established that 
an excited nuclear system (with mass A) before its disintegration may reach 
a big freeze-out volume $\approx 4/3 \pi r_0^{3} A$. In this volume the 
system can be considered in thermal equilibrium and live 
for a short time ($\sim$100 fm/c). 
We assume that the coalescence should be a rather fast process 
which happens when particles leave the interaction zone and the rate of the 
secondary interactions decreased considerably. From the UrQMD model 
calculation for 20 A GeV we evaluate this time as $\loo$50 fm/c for big 
targets and projectiles \cite{Bot11}. That is usually smaller than the 
decay time of nuclear systems in both multifragmentation and 
evaporation/fission processes: It is therefore naturally that such a 
coalescence  prescription 
may introduce an excitation energy in the clusters, which can 
decay afterwards. 
As our analysis shows, the criterion in the coordinate space correlates 
with the velocity criterion. Namely, when the projectile and 
target are well separated at later times of the reaction, the proximity 
in the velocity space means the proximity in the coordinate space too. 
This correlation appears naturally within the potential capture criterion 
\cite{Bot11}. 

Another important development of our coalescence procedure is that we assign 
the primary nucleons of initial nuclei to the residual nuclei if they did 
not interact with any particle during the collision. In some approaches 
these residues are formed by default \cite{Ton83,Bot11}. However, in the 
transport models used here (UrQMD, HSD) these nucleons preserve initialized 
momenta. The stability of the initial nuclei is not pursued in this case, 
since 
low-energy interactions inside nucleus are not precisely determined. 
We think it is a very good approximation to combine such nucleons into a 
residual cluster, especially for peripheral collisions, because of their 
initial proximity in momentum and coordinate space. We have checked that 
it is not effective to resolve the residues' problem by simply increasing the 
coalescence parameters within our procedure, since we would enforce 
artificially the formation of the lightest clusters too. In addition we 
verify the assignment of these nucleons to the residues 
by controlling their rapidities $y$: Most nucleons deviate by less than 
$\pm 0.267$ from the rapidities of the target and projectile. (This range 
$\Delta y=0.267$ is associated with the Fermi momenta of nucleons in the 
initial nucleus \cite{Bot11}.) 
As we know from the production of normal fragments 
\cite{Xi97,Sch01,Ogu11}, the realistic description of residues is important: 
According to the general picture of these reactions \cite{Bot07}, 
many hyperons can be captured by a sufficiently large piece of excited 
spectator matter, leading to the formation of hot hypermatter, which in the 
following 
decay via evaporation, Fermi-break-up, fission, or multifragmentation.


\section{Rapidity and mass distributions of fragments and hyper-fragments}

\begin{figure}[t]
\vspace{-3mm}
\includegraphics[width=0.5\textwidth]{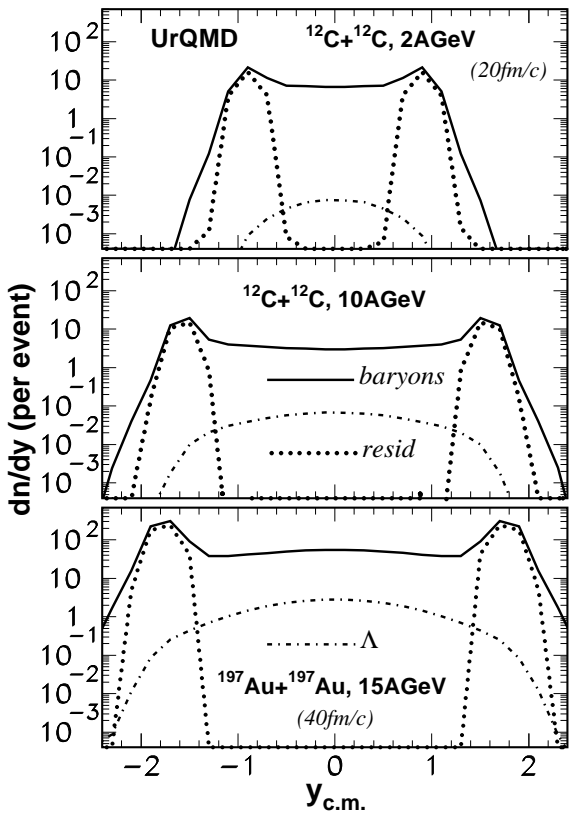}
\caption{\small{ 
UrQMD calculations for rapidity distributions of all baryons (solid lines), 
non-interacting spectator 
nucleons (dotted lines) and $\Lambda$ hyperons (dot-dashed lines), 
produced in collisions of carbon and gold beams with all impact parameters, 
normalized per inelastic event. The projectile energies and the times 
after the maximum overlap between the target and projectile 
are shown in the panels. 
}}
\label{fig1}
\end{figure}

For our analysis we have selected both small and heavy colliding nuclei, 
however, symmetric systems. For carbon projectile and targets we 
have performed calculations for different energies which are relevant 
for GSI/FAIR facility 
\cite{aumann,frs}. 
The reason is that the light hypernuclei can be quite easily identified 
in future experiments, and that such experiments are already planned 
\cite{Rap13}. The reactions with gold nuclei are added to generalize 
our conclusions for the production of heavy fragments at high beam energy. 
Such energies are easily reached with available accelerators (in 
particular, RHIC) and our predictions can help to prepare measurements 
of both light and heavy hypernuclei at all possible rapidities. We 
generated from $10^{4}$ to $10^{6}$ inelastic events for each energy 
while integrating over 
all impact parameters ('minimal' bias calculations). 

For collisions of light (carbon) nuclei we stop our transport 
calculations at the time moment of 20 fm/c after the maximum overlap between 
the target and projectile has been reached. 
We have checked that at later time cuts the number of produced particles 
and their momenta change very little, 
since they are far separated from each other. The result 
of the coalescence process remains nearly the same if we increase 
this time scale. For the gold nuclei the corresponding time 
was taken as 40 fm/c, since they are larger. 

In Fig.~1 we show the UrQMD results of the total rapidity 
distributions of all baryons and  $\Lambda$ 
hyperons produced in the collisions of a carbon projectile and target 
with laboratory beam energies of 2 and 10 GeV per nucleon, and for a gold 
on gold system at 15 GeV per nucleon. We have specially 
separated the remaining spectator nucleons which participate mostly in 
producing fragments and hyper-fragments in the target and projectile region. 
The rapidity distributions obtained with HSD model have practically the 
same form. One can see wide baryon and hyperon distributions which cover 
the whole rapidity range of the reaction. The peaks at projectile 
and target rapidities do mainly consist of the spectator nucleons which did 
not interact and which form the residues. All these particles are both the 
output 
of the transport models and the input for the coalescence approach leading 
to formation of nuclear and hyper-nuclear matter. Large excited pieces of 
hyper-matter can be produced by the capture of $\Lambda$ hyperons within 
the nuclear residues, as demonstrated also in Ref.~\cite{Bot11}. 
Similar to nuclear reactions without the involvement of strange particles, 
we expect that these hyper-residues 
will be excited and decay afterwards \cite{Bot07} producing final 
nuclei and hypernuclei. 
However, the formation of light hyper-clusters can take place at all 
rapidities. This is an advantage of the coalescence procedure as it 
can account for all these phenomena on the same footing, and, therefore, 
systematic comparisons can be performed. 

\begin{figure}[t]
\vspace{-3mm}
\hspace{-5mm}
\includegraphics[width=0.55\textwidth]{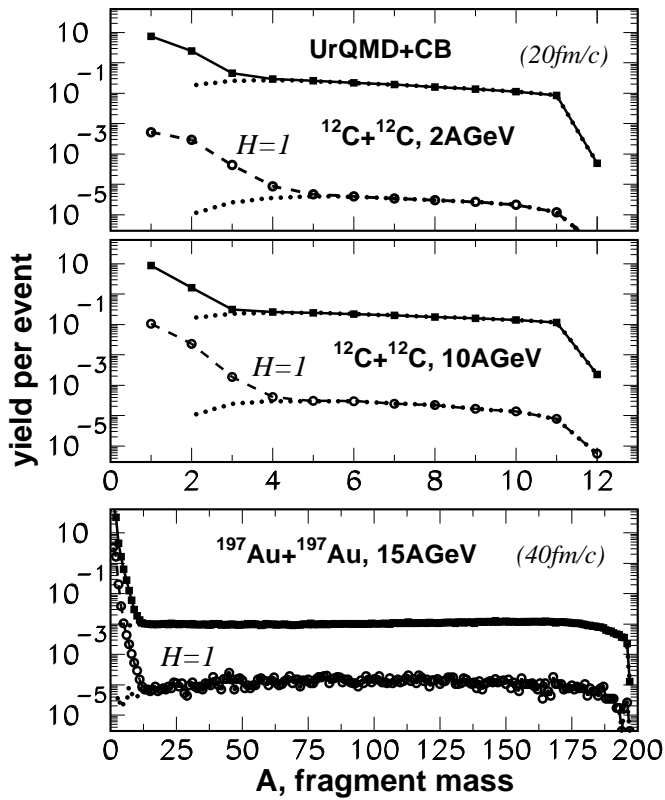}
\caption{\small{ 
Yields (per one inelastic event) of normal fragments (solid lines with 
squares) and 
hyper-fragments with one captured $\Lambda$ (notation H=1, 
dashed lines with circles) 
versus their mass number (A) in reactions 
induced by carbon and gold collisions. 
The dotted lines present the corresponding fragments originated from the 
spectator residues. 
The calculations are performed 
within the hybrid UrQMD plus CB model, with the coalescence parameter 
$v_{c} = 0.22c$, and integration over all impact parameters. The projectile 
lab energies and the transition times from UrQMD to CB are shown in 
panels. 
}}
\label{fig2}
\end{figure}

The total mass yields of the normal fragments and hyper-fragments (with one 
bound $\Lambda$) are 
shown in Fig.~2. The coalescence of baryons (the CB model) was applied after 
the UrQMD, for the reactions demonstrated in Fig.~1. The yields are normalized 
per one inelastic event. However, one should take into account that only 
events with production of hyperons have been analysed in this case. For 
this reason there is no characteristic increase of the yield of normal 
fragments with masses around the projectile/target mass, which are caused 
by very peripheral collisions. The explanation of this behaviour was 
already suggested in Ref.~\cite{Bot11}: 
The production of hyperons needs usually many particle collisions leading to 
a considerable emission of fast nucleons from the residues. 
The HSD+CB calculations show similar distributions. 

One can see that the production of fragments of all sizes is possible. As 
expected the yield of conventional fragments is by few orders of magnitude 
higher than the 
yield of hyper-fragments. Nevertheless, the production of hyper-fragments 
is sufficient to be experimentally measured (see also \cite{Bot13}). 
It is a natural result of the coalescence that the yield of the lightest 
hyper-fragments is dominating. However, the capture of hyperons by residues 
saturates the yield for large masses and leads to 
abundant production of heavy hyper-fragments. 
Within this approach one can see clearly that nearly all normal fragments 
and hyper-fragments with $A>3-4$ in the carbon collisions, and 
with $A>10$ in the gold collisions originate from the capture of $\Lambda$ 
hyperons by spectator residues (dotted lines). 
As was mentioned we believe that these hyper-fragments represent excited 
pieces of hyper-matter whose evolution can be calculated with statistical 
models \cite{Bot07, Buy13}. The excitation energy of such primary fragments 
can also be evaluated from the analysis of experimental data 
\cite{Bon95,Xi97,Sch01,Ogu11}. 

For this calculation we have used the coalescence parameter 
$v_{c} = 0.22c$ in order to take into account the higher velocities possible  
in big clusters formed by the residues. It is also consistent with the 
values obtained in our previous analysis in Ref.~\cite{Bot11}. Decreasing 
$v_{c}$ leads to a smaller yield of hyper-fragments, without changing 
the form of their distribution. In this case the yield of light normal 
fragments is reduced while residues are hardly affected.

\begin{figure}[t]
\includegraphics[width=0.5\textwidth]{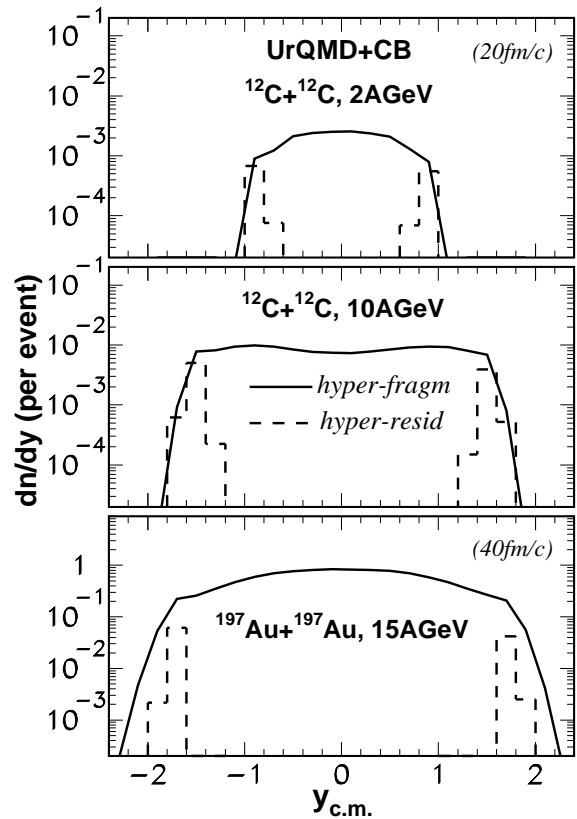}
\caption{\small{ 
Rapidity distributions (in the center of mass system, $y_{c.m.}$) 
of produced hyper-fragments (solid lines) 
and hyper-residues (dashed lines) calculated within the UrQMD plus 
CB model. 
The reactions, parameters and other notations are as in Fig.~2. 
}}
\label{fig3}
\end{figure}

\begin{figure}[t]
\includegraphics[width=0.5\textwidth]{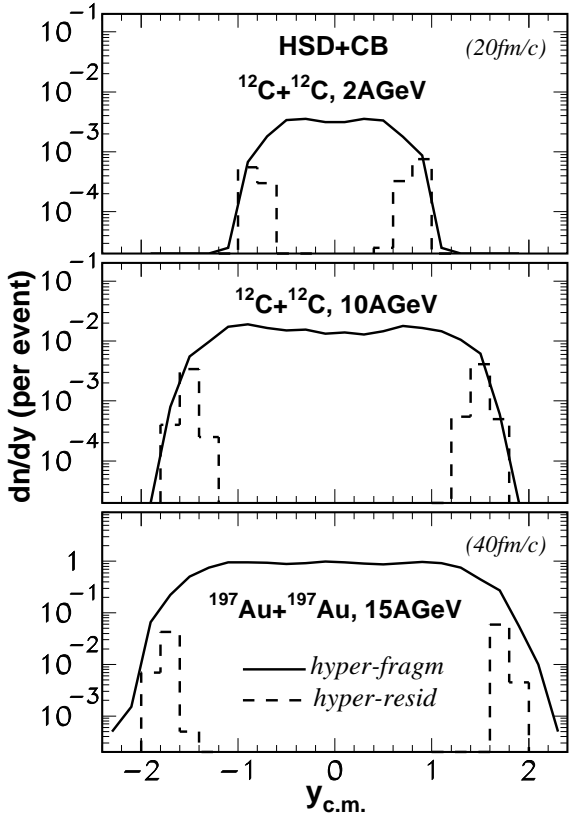}
\caption{\small{
The same as in Fig.~3 but for calculations within the HSD plus 
CB model. 
}}
\label{fig4}
\end{figure}

To complement the analysis of the fragment masses we provide information 
about the velocities of all produced hyper-fragments. 
Here and in the following figures we consider the hyper-fragments and 
hyper-residues with mass numbers $A>2$. Their total rapidity 
distributions are demonstrated in Figs.~3 and 4, for UrQMD+CB and HSD+CB 
calculations respectively. The hyper-fragments can be produced at any 
rapidity available for hyperons in the reaction (solid lines). However, 
as seen from the figures the big fragments, which can come only from the 
residues, are concentrated 
around the target and projectile rapidity (dashed lines). The small fragments 
formed after the coalescence of fast baryons can populate the midrapidity 
region also. 
As can be seen from a comparison of Fig.~3 and Fig.~4 the shape of the 
obtained 
distributions do not change with the employed transport models, 
UrQMD and HSD. However, the yields of light hyper-fragments are slightly 
larger in the HSD case. 
It is instructive that in the carbon collisions the hyper-residues are 
responsible for producing nearly all hyper-fragments in their kinematic 
regions. In the gold case, many new particles are produced in this region, 
therefore, besides big hyper-residues additional light hypernuclei can be 
formed too.

\begin{figure}[t]
\includegraphics[width=0.5\textwidth]{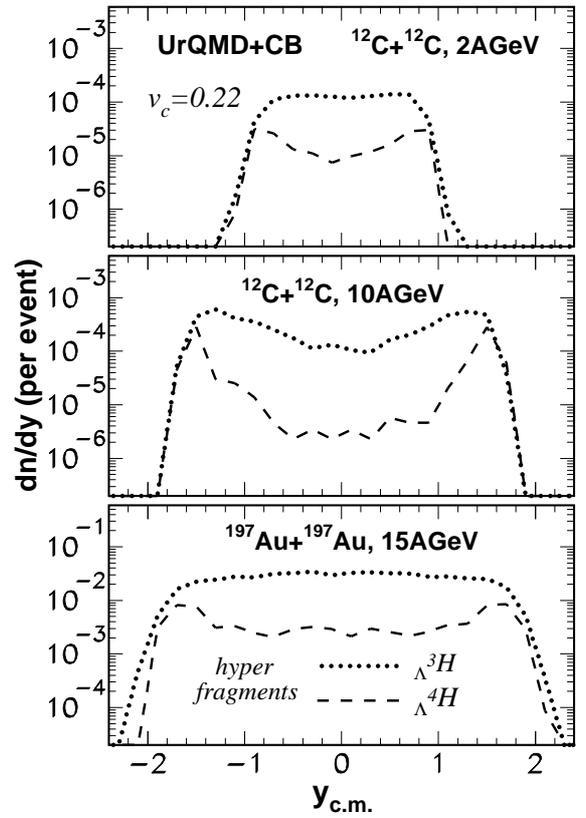}
\caption{\small{
Rapidity distributions of produced $^{3}_{\Lambda}$H (dotted lines) and 
$^{4}_{\Lambda}$H (dashed lines) hyper-fragments in reactions as in 
Fig.~3. The UrQMD and CB calculations are with the coalescent parameter 
$v_{c} = 0.22c$
}}
\label{fig5}
\end{figure}

\begin{figure}[t]
\includegraphics[width=0.5\textwidth]{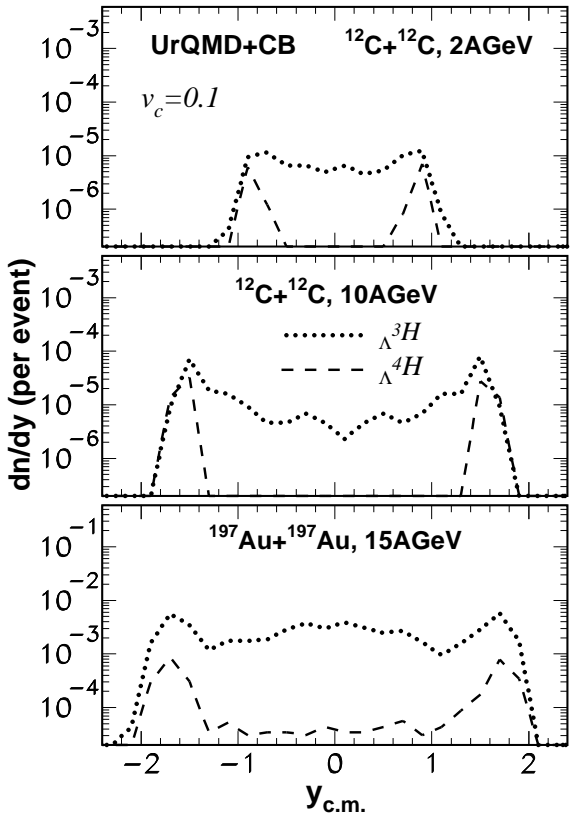}
\caption{\small{ 
The same as in Fig.~5 but with the coalescent parameter $v_{c} = 0.1c$
}}
\label{fig6}
\end{figure}

The light hypernuclei $^{3}_{\Lambda}$H and $^{4}_{\Lambda}$H are specially 
interesting: They can be easily identified by their decay into 
$\pi^{-}$ and $^{3}$He, and into $\pi^{-}$ and $^{4}$He, respectively. 
These correlations have been observed already in many heavy-ion experiments 
at high energies \cite{star,alice,ygma-nufra,saito-new,Rap13b}. 
Such hypernuclei can serve as indicators for the production of hyper-matter. 

In Figs.~5 and 6 we show the rapidity distributions of these light hypernuclei 
produced in the same reactions. The simulations are performed within 
the UrQMD and CB models. As before, the coalescence parameter 
$v_{c} = 0.22c$ has been used for the calculations shown in Fig.~5. 
For comparison, the results obtained with a smaller parameter $v_{c} = 0.1c$ 
are presented in Fig.~6. The latter may be more adequate for these 
small nuclei, since previously the yields of normal small clusters have 
been well described with a such low coalescence parameter \cite{Ste12,Ton83}. 
In this case the fragments can be treated already as nuclei in final state 
without secondary de-excitation, since the later one is mainly relevant 
for big residues. 

One can see an interesting behaviour: The $^{3}_{\Lambda}$H nuclei are 
essentially formed over all rapidities. It is obvious, that the 
production of the clusters is smaller at low 
coalescent parameters (compare Figs.~5 and 6).
At low $v_{c}$, however, 
the fragments are more grouped at the target and projectile 
rapidities. This concentration is more evident for lager nuclei -- 
$^{4}_{\Lambda}$H. This is the consequence of the applied coalescent 
mechanism: 
If the velocity space is reduced the large clusters are more efficient 
in the capture of hyperons because of the larger coordinate space. 

With increasing energy the fraction of nuclei around 
residues increases, since more particles are produced in this region 
as a result of secondary interactions. Whereas particles 
originating from midrapidity have higher energy and they 
are more separated in the phase space. Therefore, despite of the general 
increase the number of such particles, the total number of clusters 
may not increase.

\begin{figure}[t]
\includegraphics[width=0.5\textwidth]{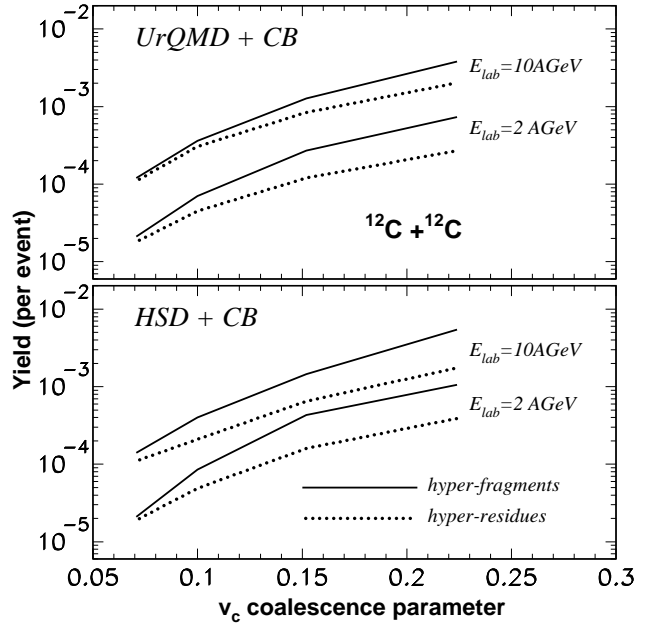}
\caption{\small{
Yields of all produced hyper-fragments (solid lines) and hyper-residues 
(dotted lines) versus the coalescent parameter $v_{c}$, as calculated 
within the UrQMD and CB model (top panel) and HSD and CB model (bottom 
panel). The reaction is the carbon on carbon collisions (integrated 
over all impact parameters) with the projectile lab energies of 
2 and 10 GeV per nucleon, see notations by the lines. 
}}
\label{fig7}
\end{figure}

\begin{figure}[t]
\includegraphics[width=0.5\textwidth]{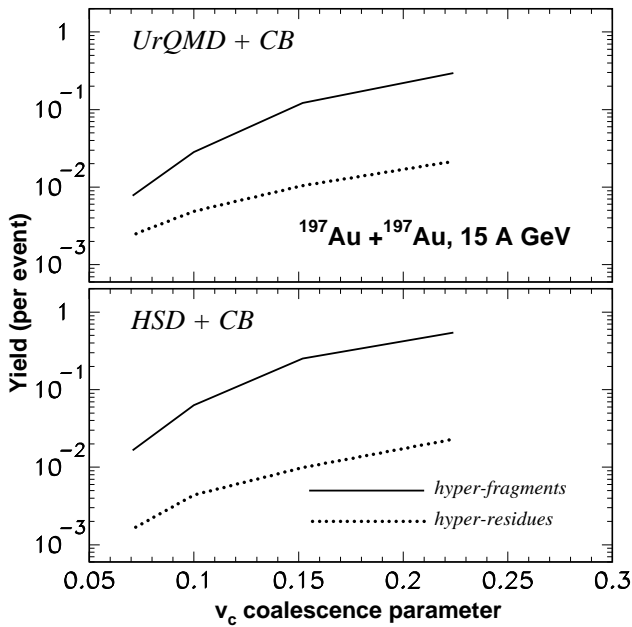}
\caption{\small{
The same as in Fig.~7 but in the reaction of the gold on 
gold collision at 15 GeV per nucleon. 
}}
\label{fig8}
\end{figure}

The dependence of the results on the coalescence parameter was specially 
investigated in these reactions, since it is related to the main uncertainty 
of the predictions. Figs.~7 and 8 show the total yields of hyper-fragments 
and hyper-residues in collisions of carbon on carbon, and gold on gold, 
respectively. As usual, the yields are normalized per inelastic event and 
integrated over all impact parameters. One can see that the UrQMD as well as 
the HSD model gives similar results. As expected all yields increase with 
$v_{c}$. However, at small $v_{c}$ the capture of hyperons take place on 
residual nuclei predominantly. The big residues cover a large coordinate 
space region that becomes important for this mechanism in the case of reducing 
the momentum space. It is also essential that the secondary interactions 
which contribute considerably to the formation of hyperons with relatively low 
momenta happen mostly in the residue region. On the other hand, 
primary interactions leading to the production of high-energy hyperons take 
place in the central (midrapidity) region. In this case particles are far 
from each other in momentum space, therefore, in order to construct 
a cluster a larger $v_{c}$ is required. 

In the gold collisions the difference between the calculations with 
UrQMD and HSD for light hyper-fragments is more prominent, by about a factor 2 
(see Fig.~8). 
Within the models this is related to the treatment of the secondary 
interactions and HSD leads to an enlarged production of these 
hyper-fragments. However, the values predicted by both models look quite 
reasonable, and they can be checked by analysing experimental data. 
In turn, the possible variation of the predictions is 
important for planing future measurements. 

\begin{figure}[tbh]
\includegraphics[width=0.5\textwidth]{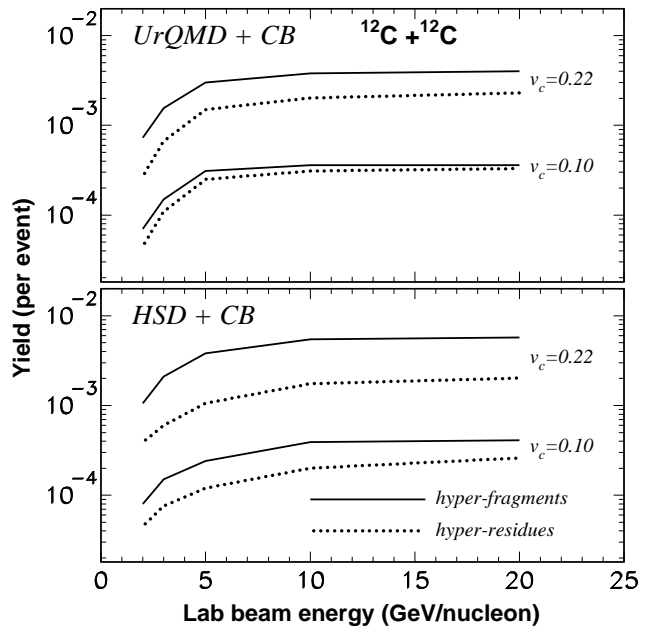}
\caption{\small{ 
Yields of all produced hyper-fragments (solid lines) and hyper-residues 
(dotted lines) versus the beam energy in the carbon on carbon collisions 
for all impact parameters, as calculated 
within the UrQMD and CB model (top panel) and HSD and CB model (bottom 
panel). The coalescent parameters $v_{c}$ are given in the panels. 
}}
\label{fig9}
\end{figure}

Since increasing the beam energy results in a larger number of produced 
hyperons, the yields of hyper-fragments may increase too. This is clearly seen 
in Fig.~7. More details are shown in Fig.~9 for carbon collisions 
for a wide range of beam energies. 
There is a saturation of the yields of hyper-fragments, both light and 
heavy ones, at energies higher 
than 5--10 GeV per nucleon. This effect is found for both models and for all 
coalescent parameters. Depending on $v_{c}$ this saturation 
happens at a different level. By comparing these results with the 
previous ones which were obtained with the DCM and the capture of hyperons 
by the nuclear 
potential (see Fig.~2 in Ref.~\cite{Bot13}) we note that they will be 
similar if we take the values of $v_{c}$ in-between the ones shown now in 
Fig.~9 (i.e., in-between 0.1 and 0.22). 
The uncertainty obtained with this parameter should be clarified by a 
comparison with 
experimental data and with more sophisticated theory calculations. 
Because of the saturation of the yield at high energies 
the experimental hypernuclear studies can be pursued at the 
accelerators of moderate relativistic energies, above the threshold 
($\goo$1.6 A GeV). 

Generally, this combination of the transport and coalescence models can 
be used for analysis of yields of non-strange fragments too. This can give 
an additional insight into the reaction mechanism. Besides light fragments 
which were already tested \cite{Ste12,Ton83}, the intermediate and large 
fragments are also of considerable interest, e.g., see the ALADIN data 
\cite{Xi97,Ogu11}. As mentioned, a detailed comparison may require a 
connection with the secondary 
de-excitation processes.


\section{Conclusion}

We conclude that relativistic heavy-ion reactions are a very promising 
source of hyper-matter and hyper-fragments. A large amount of hypernuclei 
of all masses 
can be produced. Their properties can also be investigated taking into 
account the advantages of relativistic velocities, e.g., for the life-time 
and correlation measurements. 

The well established UrQMD and HSD transport models have been used in order to 
describe the strangeness and hyperon formation. They give a quite reliable 
picture of the reactions and they also consistent with other dynamical 
approaches 
(e.g., DCM) used by us previously. The interaction of hyperons with nucleons 
leads to their capture and to the formation of hyper-matter. We describe this 
process within a generalized 
coalescence model. The coalescence of baryons is consistent with 
the hyperon capture in a potential well of large nuclear residues, and 
the coalescence parameters are expected to be of the same order as for 
normal fragments. 
This procedure gives a possibility to consider the formation of light 
hypernuclei on the same footing. We demonstrate that big hyper-fragments 
are mostly produced from the spectator residues, while the light ones 
can be formed at all rapidities. We expect, however, that some 
large species of hypermatter will be excited, and decay afterwards with 
production final hypernuclei and normal nuclei, as in usual fragmentation 
and multifragmentation reactions. Such a mechanism should allow the 
investigation of possible 
phase transitions in hypermatter with statistical models describing the 
secondary disintegration. 

By summing up the results obtained with various models we note that the 
production of hyper-fragments in relativistic heavy-ion collisions is 
universal 
and well established theoretically. It is very instructive to investigate 
the whole reaction mechanism by measuring simultaneously big hypernuclei 
originating from the residues and light hypernuclei which can be formed 
in the hot midrapidity region. The saturation of all yields takes place 
at the beam energies higher than 5--10 GeV per nucleon. This opens the 
possibility to study hypernuclei at GSI/FAIR (Darmstadt), Nuclotron/NICA 
(Dubna), 
RHIC (Brookhaven), HIAF (Lanzhou) and other heavy-ion accelerators of 
moderate relativistic energies.
In the following we plan to analyze theoretically the formation of 
multi-hyperon nuclei, which can be abundantly produced in these reactions. 
Another promising opportunity would be to study unstable (resonance) states of 
hypernuclei via particle correlations. In addition, 
exotic hypernuclei may be formed and investigated in the secondary 
evaporation, fission, and multifragmentation--like processes. 
Comparing these theoretical predictions with future experiments may provide 
new information on the $YN$ and $YY$ interaction at low energies, as 
well as about properties of hyper-matter.

\section{Acknowledgments}

This work was supported by the the GSI Helmholtzzentrum f\"ur 
Schwerionenforschung and Hessian initiative for excellence 
(LOEWE) through the Helmholtz International Center for FAIR (HIC for FAIR)
We also acknowledge the support by the Research Infrastructure Integrating 
Activity Study of Strongly Interacting Matter HadronPhysics3 under the 7th 
Framework Programme of EU (SPHERE network). 
The computational resources were provided by the LOEWE Frankfurt 
Center for Scientific Computing (LOEWE-CSC). 




			\end{document}